\documentclass[sigconf]{acmart}
\AtBeginDocument{%
  \providecommand\BibTeX{{%
    \normalfont B\kern-0.5em{\scshape i\kern-0.25em b}\kern-0.8em\TeX}}}

\usepackage{wrapfig}
\usepackage{makecell}
\usepackage[ruled,vlined]{algorithm2e}
\usepackage{graphicx}
\usepackage{caption}
\usepackage[htt]{hyphenat}
\usepackage{enumitem}
\setdescription{topsep=1.5ex, leftmargin=0ex, itemsep=1.5ex}
\setitemize{topsep=1.5ex, leftmargin=3ex, itemsep=1.5ex}
\setenumerate{topsep=1.5ex, leftmargin=3ex, itemsep=1.5ex}

\usepackage{xcolor} 
\newcommand{\tablegrey}{black!10} 

\usepackage{colortbl}
\usepackage{etoolbox}
\usepackage{booktabs}
\newcommand{\greyrule}{\arrayrulecolor{\tablegrey}\midrule\arrayrulecolor{black}} 
\usepackage{subcaption}

\makeatletter
\newcommand{\removelatexerror}{\let\@latex@error\@gobble}
\makeatother

\copyrightyear{2023}
\acmYear{2023}
\setcopyright{rightsretained}
\acmConference[AISec '23]{Proceedings of the 16th ACM Workshop on Artificial Intelligence and Security}{November 30, 2023}{Copenhagen, Denmark}
\acmBooktitle{Proceedings of the 16th ACM Workshop on Artificial Intelligence and Security (AISec '23), November 30, 2023, Copenhagen, Denmark}\acmDOI{10.1145/3605764.3623986}
\acmISBN{979-8-4007-0260-0/23/11}
\makeatletter
\gdef\@copyrightpermission{
  \begin{minipage}{0.3\columnwidth}
   \href{https://creativecommons.org/licenses/by/4.0/}{\includegraphics[width=0.90\textwidth]{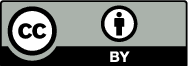}}
  \end{minipage}\hfill
  \begin{minipage}{0.7\columnwidth}
   \href{https://creativecommons.org/licenses/by/4.0/}{This work is licensed under a Creative Commons Attribution International 4.0 License.}
  \end{minipage}
  \vspace{5pt}
}
\makeatother
\begin{document}

\title{Canaries and Whistles: Resilient Drone Communication Networks with (or without) Deep Reinforcement Learning}

\author{Chris Hicks}
\email{c.hicks@turing.ac.uk}
\affiliation{%
  \institution{The Alan Turing Institute}
  \country{}
}

\author{Vasilis Mavroudis}
\email{vmavroudis@turing.ac.uk}
\affiliation{%
  \institution{The Alan Turing Institute}
   \country{}
}

\author{Myles Foley}
\email{m.foley20@imperial.ac.uk}
\affiliation{%
  \institution{Imperial College London}
   \country{}
}

\author{Thomas Davies}
\email{tdavies@turing.ac.uk}
\affiliation{%
  \institution{The Alan Turing Institute}
   \country{}
}

\author{Kate Highnam}
\email{k.highnam19@imperial.ac.uk}
\affiliation{%
  \institution{The Alan Turing Institute}
   \country{}
}

\author{Tim Watson}
\email{tim.watson@turing.ac.uk}
\affiliation{%
  \institution{The Alan Turing Institute}
   \country{}
}

\renewcommand{\shortauthors}{Chris Hicks et al.}
\hyphenation{Petting-Zoo-Parallel-Wrapper}
\hyphenation{Remove-Other-Sessions}

\begin{abstract}
Communication networks able to withstand hostile environments are critically important for disaster relief operations. In this paper, we consider a challenging scenario where drones have been compromised in the supply chain, during their manufacture, and harbour malicious software capable of wide-ranging and infectious disruption. We investigate multi-agent deep reinforcement learning as a tool for learning defensive strategies that maximise communications bandwidth despite continual adversarial interference. Using a public challenge for learning network resilience strategies, we propose a state-of-the-art expert technique and study its superiority over deep reinforcement learning agents. Correspondingly, we identify three specific methods for improving the performance of our learning-based agents: (1) ensuring each observation contains the necessary information, (2) using expert agents to provide a curriculum for learning, and (3) paying close attention to reward. We apply our methods and present a new mixed strategy enabling expert and learning-based agents to work together and improve on all prior results. 
\end{abstract}



\maketitle

\section{Introduction}
Drones, and other types of unmanned aerial vehicle, are increasingly utilised for disaster relief efforts where they can help to co-ordinate efforts on the ground by providing real-time surveillance, ad-hoc communication networks, and the delivery of supplies to remote areas~\cite{glantz2020uav}. Drones are low-cost, widely available and can operate in hostile and warlike environments with degraded or inoperable infrastructure. When multiple drones are available, connecting them together wirelessly can provide an ad-hoc communication network with enhanced coverage, resilience and safety. Unfortunately, as with embedded systems at large~\cite{costin2014embeded,yu2022embedded}, numerous commercial and military drones are vulnerable to cyber attacks~\cite{ronen2017zigbeeworm, dey2018uav, mit2019djiattack, rand2020dronecyber,ce2022doublestar,mekdad2023uavsecsurvey}. Even if a drone is secure by design, malware can attack the supply chain to compromise the functionality of a specific component~\cite{belikovetsky2017drowned}. Drone vulnerabilities have been exploited to provide location tracking~\cite{kerns2014gpsspoof}, botnet attacks~\cite{reed2011skynet}, sensor tampering~\cite{mit2019djiattack,ce2022doublestar} and covert data exfiltration~\cite{valente2017threats}. Ultimately, many drone vulnerabilities give complete control to the adversary, allowing them to interfere with the drone as they please.

Given the ubiquity of software vulnerabilities in drone systems, it is essential to anticipate their compromise during operation. If drone malware could be actively impeded during operation, then despite the inevitable attacks, an ad-hoc network of drones with sufficient redundancy could still provide valuable services. Current mitigations for drone vulnerabilities include regular operating system updates, intrusion detection systems (IDS), fine-grained circuit analysis, and remote software attestation~\cite{mekdad2023uavsecsurvey}. However, these defences are not sufficient in many cases. Operating system updates often take several weeks, at best, to include patches for the latest exploits. IDS can help detect adversaries but, in addition to being computationally expensive, are usually limited to observing network traffic without context and can negatively impact network latency. Fine-grained circuit analysis can be defeated~\cite{gohil2022attrition} and remote software attestation depends, in addition to being certain that attested software is secure, on managing the multiple drone authorisations for swarm-based solutions. None of the current defences can actively impede malware from affecting drones, maintaining crucial ad-hoc communication networks for disaster relief. 

Autonomous cyber defence (ACD) is a class of solution methods, models and intelligent agents that actively respond to cyber attacks without the need for human intervention. Intelligent agents trained using reinforcement learning (RL) have, in particular, shown great potential for autonomously defending computer networks and systems from attack~\cite{standen2021cyborg,foley2022CAGE1,foley2022CAGE2}. ACD is particularly valuable when, as in the case of a disaster relief, there is a lack of human experts available to defend systems against an adversary. People on the ground dealing with the situation, emergency responders and regular civilians, are unlikely to have the skills, resources or time needed to patch vulnerable drones. 

In this paper we evaluate multi-agent RL (MARL) for defending ad-hoc communication networks against malware attacks. Using a public challenge and environment for ACD blue teaming, we first propose a state-of-the-art expert agent, \emph{Canary}. Next, we show that RL applied in the standard setting yields unsatisfactory results. We identify a framework for bridging the gap in performance, apply our methods and demonstrate the potential of learning-based policies in actively defending ad-hoc drone networks. 

\section{Background and Motivation}



\subsection{Reinforcement Learning (RL)}
RL is about learning how to interact with an unknown environment to maximise a numerical reward signal. RL is characterised by \emph{trial-and-error}, in which the learner discovers through repeated interaction which actions lead to success, and \emph{delayed reward}, whereby actions may affect rewards far in the future. An important property of RL is that goals are defined only by specifying the reward function. There is no need to define exactly how to reach the goal. RL provides a mechanism to distribute the long-term rewards of goal accomplishment among the many actions that contributed to success. 


\begin{figure}[h]
    \centering
    \includegraphics[width=1.0\columnwidth]{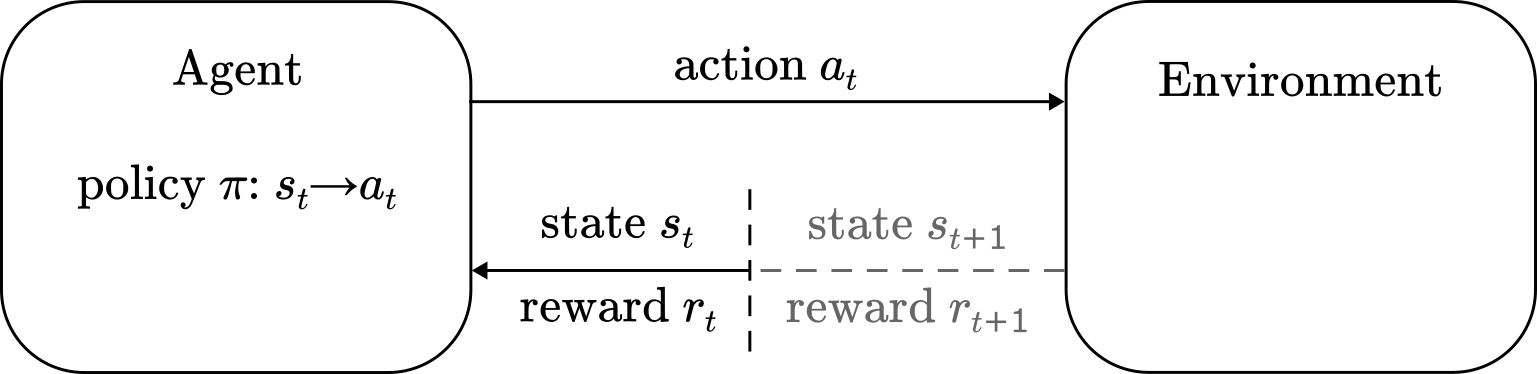}
    \caption{The sequential agent-environment interaction formalisation of RL.}
    \label{fig:rlformalisation}
\end{figure}

As shown in Figure \ref{fig:rlformalisation}, RL is characterised by a sequential interaction between an \emph{agent} and an \emph{environment}. At each time step $t$ the agent observes the state $s_t$, the reward $r_t$, and chooses an action $a_t$. The environment is formalised as an incompletely-known Markov decision process (MDP) i.e., future states $s_{t+1}$ are independent of past states $s_0,\ldots, s_{t-1}$ given the present state $s_t$. Learning from interaction takes place over episodes, each comprising multiple time steps. Where $H$ is the maximum number of time steps in an episode, the agent uses trajectories $\tau = \{(s_0,a_0,r_0), \ldots, (s_H,a_H,r_H)\}$ to learn a policy $\pi$ which is a map from the state space $S$ to the action space $A$. 


Recently, RL algorithms have demonstrated human and super-human level performances including beating world champions at Chess and Go~\cite{silver2016go,silver2018alphazero}, mastering Stratego~\cite{perolat2022stratego} and discovering faster sorting algorithms~\cite{mankowitz2023sort}. These advances have been made possible by DRL, in which the policy of an RL agent is represented by a deep neural network~\cite{mnih2015dqn,schulman2017ppo}. DRL enables agents to handle large state spaces by generalising from the trajectories seen during training to rarely or never-before seen states. In addition, neural function approximation relaxes the MDP environment requirement of early tabular RL algorithms~\cite{watkins1992qlearning} to a partially observable MDP (POMDP)~\cite{sutton2018rlbible}. This means that DRL agents can learn policies in non-Markovian environments by parameterising only the Markovian variables~\cite{whitehead1995hidden}. Even so, performance is diminished in proportion to the lack of Markovian variables and state representations must be carefully constructed to include all of the information needed to properly inform each action.  

When RL is extended to include multiple agents it is termed multi-agent RL (MARL). MARL enables a range of cooperative, competitive and mixed strategies to emerge between decentralised autonomous agents~\cite{baker2020emergent}. In large problem spaces MARL also permits divide-and-conquer strategies based on specialised agents handling specific sub-problems. Despite the advantages of MARL, it is also considerably more challenging than the single-agent setting. Multiple agents learning simultaneously present a much more difficult non-stationary learning problem~\cite{whitehead1995hidden}. In addition, assigning long-term rewards to specific actions is greatly frustrated when multiple agents contribute to, and detract from, goal accomplishment. The more agents coexist in a MARL environment, the more intractable it is from a learning standpoint~\cite{yang2018marlmean}.

\subsection{Curriculum Learning}
Curriculum learning (CL)~\cite{bengio2009curriculum} is an approach that can help agents learn more effectively by first training on simpler tasks and then gradually increasing the complexity. The main idea is to continuously adjust the difficulty of the task to just beyond the current capabilities of each agent, allowing them to learn new skills gradually. CL is a useful strategy for overcoming the specific difficulties of MARL, including non-stationary learning, and even emerges implicitly in certain cooperative and competitive settings~\cite{silver2016go,wang2020coopcurric,baker2020emergent}.


\subsection{Autonomous Cyber Defence}
Motivated by the growing shortfall in cyber skills, scale and speed of response that is required to defend modern digital infrastructure from cyber attacks, ACD concerns the development of autonomous agents that actively defend computer networks and systems without the need for human intervention. Once threats have been detected and identified, ACD systems take actions to protect against, respond to, and recover from attacks. ACD systems can also deploy countermeasures including decoys, canaries and honeynets to gain defensive advantages against the adversary~\cite{lohn2023acdpolicy}. The most recent advances in ACD have emerged from the application of state-of-the-art DRL methods~\cite{hanrl4sdrsec,hammar2020finding,foley2022CAGE1,campbell2023curricrl4sec,foley2022CAGE2,cage2aicd}.

An important aspect of the ACD landscape are cyber simulator environments, i.e, AI gyms~\cite{brockman2016openai}, that enable autonomous agents to be trained without the scaling limitations of real systems and networks~\cite{burke2022acd}. At least 16 different ACD simulation environments have been described in the literature, however only Cyber Operations Research Gym (CybORG)~\cite{standen2021cyborg, cage_cyborg_2022} is both open source and designed specifically for training defensive RL-based agents~\cite{vyas2023acd}. CybORG is an AI gym, and research platform, providing a flexible scenario-driven environment backed by both simulated and emulated (i.e., Amazon Web Services) networks. A number of offensive red agents are included in CybORG to allow for benchmarking defensive strategies. Significantly, CybORG has hosted three competitive public challenges which have motivated the development of autonomous agents by the wider academic community~\cite{foley2022CAGE1,foley2022CAGE2,applebaum2022tabqcage,wolk2022cage}. 

Despite the apparent need for MARL approaches to scale ACD solutions to the intractably large observation spaces of real and internet-scale computer networks, only two cyber simulation environments currently support MARL~\cite{kunz2022multiagent,cage_cyborg_2022}. Of these, only the latest CybORG challenge~\cite{cage_challenge_3_announcement} is defence focused. 



\section{Attacker Model}
We consider a strong, yet realistic~\cite{dod2022supplychains}, adversary who has compromised the supply chain of a drone manufacturer. The adversary has covertly placed malware on the firmware installed onto every drone. The malware lays dormant until such a time as it is activated by the adversary, whereupon it launches a variety of harmful attacks including passively listening in on communication and actively compromising neighbouring drones. These drones have already been deployed and cannot quickly be replaced. The drones must be used as-is, despite having compromised firmware, to provide a temporary ad-hoc communication network for people on the ground. Whilst being unable to fully remove the malware, we will investigate the possibility of providing an active defence such that the network can still provide communications bandwidth.

\section{Environment} \label{sec:description}

To study resilient ad-hoc communication networks from the perspective of malware compromised drones, we use the CybORG AI gym and its latest public competition~\cite{cage_challenge_3_announcement}. The third CAGE challenge, hereafter referred to as ``the CAGE challenge'' for brevity, tasks competitors with developing an autonomous defensive capability for an ad-hoc network of drones. The drones have been compromised during their manufacturing process and harbour malware that cannot easily be removed. Despite this, a communication network is desperately needed and the drones must be put into service. The CAGE challenge provides a MARL research environment for determining to what extent communication bandwidth can be maintained, by an ad-hoc network of drones, in the face of a disruptive malware attack.

\begin{figure}[b]
    \centering
    \includegraphics[width=0.8\columnwidth]{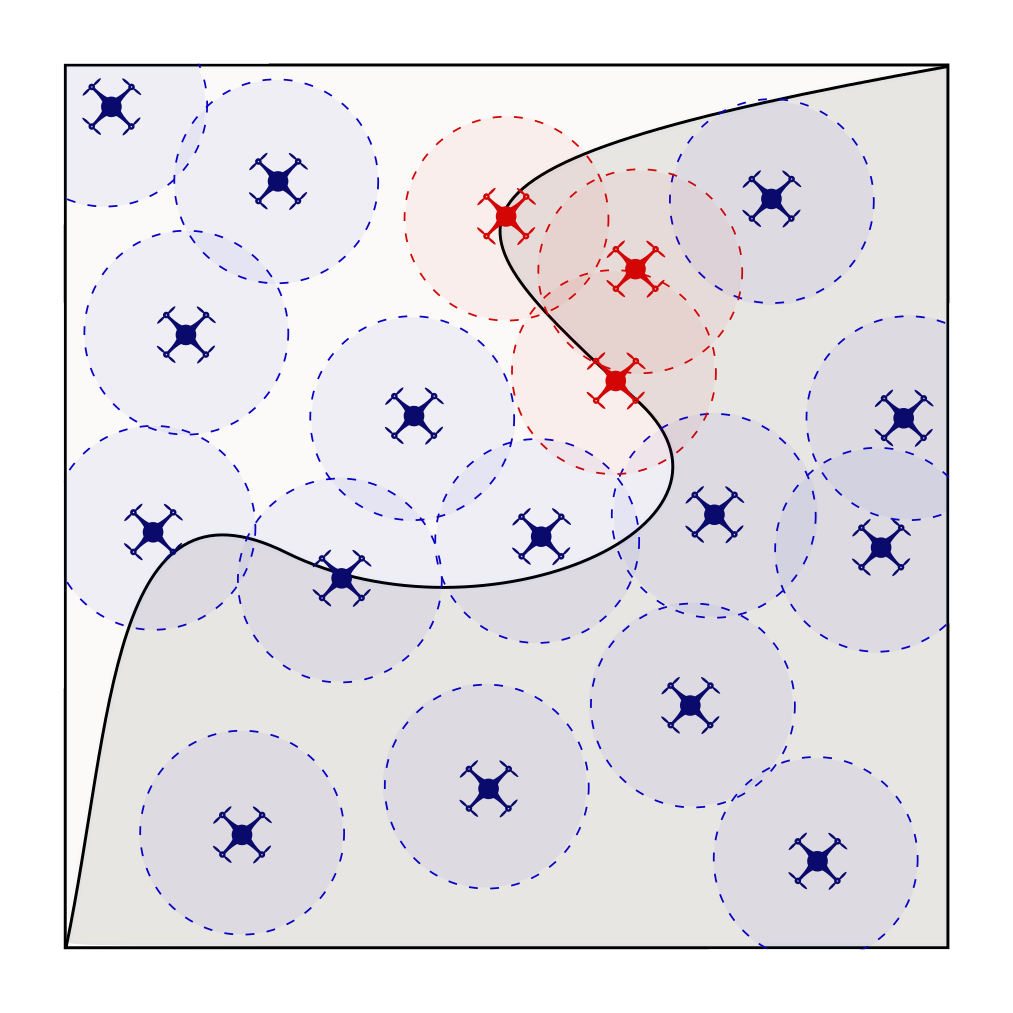}
    \caption{The CAGE Challenge Scenario. Blue and red drones are controlled by the defensive and offensive teams, respectively. Shaded areas illustrate the (not always) overlapping radii of each drone's communication range.}
    \label{fig:cc3_battlefield}
\end{figure}

The CAGE challenge uses the CybORG environment~\cite{cage_cyborg_2022} to define a specific drone swarm scenario illustrated in Figure~\ref{fig:cc3_battlefield}. The scenario comprises 18 drone hosts that are perpetually at risk from a malware that lies dormant in their firmware. The drones occupy a $100\times 100$ two-dimensional space, have a communication radius of 30, and a maximum bandwidth of 100. At discrete time steps, a defensive blue team, an offensive red team and a neutral green team take turns to accomplish their objectives. The CAGE challenge provides both red and green teams as part of the environment, positioning for researchers the task of controlling the defensive blue team. The green team has one agent for each drone and is used to simulate the demand for communication bandwidth by operatives on the ground. Meanwhile the blue and red teams compete to control the drones, meaning a total of 18 agents are active across both teams at any moment. The motion of each drone, hence the overall network topology, is controlled by a randomised swarming algorithm unaffected by any of the teams. This ensures the task of actively mitigating malware attacks using software command and control (C2) tactics can be researched as an independent variable.

The development of RL agents for this purpose is explicitly encouraged by the reward function, specified in Table \ref{table:reward_function}, that is returned at every time step through the standard AI gym interface~\cite{brockman2016openai}. The reward function is also used to evaluate and benchmark the performance of defensive agents by averaging their scores over 1000 episodes, each up to 500 steps long. The maximum that can be scored in the CAGE challenge is $0$, corresponding to flawless message delivery, and the minimum is $-9000$, equivalent to all messages failing for an entire episode. The results for 12 different approaches from 8 different teams are ranked publicly now that the challenge has officially closed~\cite{cage_challenge_3_announcement}. Our Canary protocol, which we introduce in Section~\ref{sec:canary}, currently ranks as the top performing agent\footnote{Preliminary challenge results correct as of 14th July 2023.}.

\begin{table}[t]

\centering
\begin{tabular}{p{0.29\columnwidth}p{0.39\columnwidth}}
Reward & Event  \\
\midrule
$-1$ & Comms. blocked  \\ \greyrule
$-1$ & Comms. intercepted  \\ \greyrule
$-1$ & Comms. dropped due to insufficient bandwidth  \\ \greyrule
$-1$ & No comms. route can be established  \\ \greyrule
$-(18*(t-500))$ & Complete compromise  \\ 
\bottomrule
\end{tabular}
\caption{The CAGE challenge reward function.}
\label{table:reward_function}
\end{table}




\subsection{Drone Platform}
Each drone is a simulated embedded Linux system with a wireless interface that establishes a data link with all neighbouring drones within communication range. All drones have the same firmware-level malware installed which, when activated, will passively and actively disrupt the swarm whilst also spreading to neighboring drones. To help keep drones secure, they include an Internet Protocol (IP) block list and a tool for detecting malicious network activity. The drones have 8 different actions that are selectively available to red, blue and green agents. Actions are either \emph{local}, and act on the agent's host drone, or \emph{remote}, and act on another drone in the swarm. Remote actions consume bandwidth and each drone has a maximum of 100 units to spare, per time step, before all additional actions (and messages) are dropped. To account for stochasticity in real network attack and defence scenarios, several of the actions occasionally fail. 

In addition to the communication network provided by the drone network at large, an explicit low-bandwidth broadcast channel is also provided. At every time step, each drone can broadcast a 16~bit message to all neighbouring systems within range. Since each drone only benefits from a partial observation of the environment, inter-drone communication is essential to effectively coordinate a defensive response.



    

\subsection{Green Team}\label{subsec:greenteam}
The CybORG environment uses a green team to simulate the demand for communication bandwidth. At every time step, the green agent on each drone chooses a random recipient drone from the swarm and an arbitrary message is sent, consuming 1 unit of bandwidth from each drone along the shortest route (as determined by Dijkstras algorithm~\cite{dijkstra1959note}). Importantly, noting the reward function in Table~\ref{table:reward_function}, a negative reward is expected regardless of the disruption caused by malware. Green agents cause negative reward by choosing target drones that are unreachable owing to the randomised network topology.








\subsection{Malware}\label{subsec:malware}
There is a firmware-level malware on each drone that cannot easily be removed. Once activated, occurring on a random drone with a 5\% independent probability at each time step, the malware immediately removes the blue team's agent and replaces it with a high-privilege red team agent (controlled by the malware). The activated malware is able to fully control the drone beginning the same time step. Until the drone is reclaimed by the blue team, the malware behaves according to one of 6 different strategies which are chosen, and switched between, at random.

\begin{description}
    \item [ExploitRandom.] Exploit as many neighbouring drones as possible. The malware first remotely exploits a neighbouring drone at random yielding a low-privilege red agent presence on the target. Remote exploitation is flagged as malicious 50\% of the time, independently by each host along the route), and also fails 50\% of the time. If the low-privilege red agent session has not been removed by the next time step (i.e., by the blue agent), then the agent will escalate privileges to high (and cannot fail). A high privilege red agent represents complete activation of malware on the target.

    \item [{ExploitFarthest}.] This strategy is the same as \texttt{ExploitRandom}. However, drones are targeted from farthest to nearest.

    \item [{FloodRandom}.] Consumes all available bandwidth on the target drone which is chosen at random from the swarm. This action will never fail, provided the target is reachable using the drone network (i.e., is routable and the relevant IP addresses are not blocked). This action is always flagged as malicious by drones along the route.

    \item [{FloodFarthest}.] This strategy is as \texttt{FloodRandom} however target drones are chosen from farthest to nearest.
    
    
    \item [{BlockRandom}.] A neighbouring drone within direct communication range is selected at random and added to the block list. All subsequent attempts to reach the target via the infected host will be dropped until either 1. the end of the episode, or 2. a blue agent resumes control and unblocks the relevant IP address.


    \item[Intercept.] The malware silently intercepts messages on behalf of the red team. This strategy is never flagged as malicious and does not fail. 
    
\end{description}

Every turn that the malware executes one of these strategies, there is also an approximately 10\% chance that the strategy will be swapped. 

\subsection{Blue Team}\label{subsec:blueteam}
The blue team begins each episode with high privilege control of all 18 drones. The CybORG environment provides a standard AI gym interface for training RL agents which, as in Figure~\ref{fig:rlformalisation}, includes an action space, a state space and a reward signal. An additional PettingZoo~\cite{terry2021pettingzoo} wrapper is provided which provides a standard API for MARL environments. 

\begin{description}
    \item[Action Space.] Defensive blue agents can use the following actions. The RetakeControl, BlockTraffic and AllowTraffic actions are each parameterised by the IP address of the target ($n=18$), in total providing an action space of $56$ (i.e., $1+18+18+18+1$).
\end{description}

\begin{itemize}

\item \textbf{RemoveOtherSessions.} Remove locally any low privilege red agent session from the host drone, succeeding 90\% of the time. The agent can observe whether the action succeeded on the next time step.

\item \textbf{RetakeControl.} Removes any red agents on the remote target, including high privilege, and returns control to the blue team. This action succeeds with a 75\% probability, consumes 10 units of bandwidth from each drone along the network route, and is falsely identified as malicious activity 15\% of the time.

\item \textbf{BlockTraffic.} Adds the target drone to the IP block list, blocking all subsequent traffic until either 1. the end of the episode, or 2. a blue agent unblocks the relevant IP address. 

\item \textbf{AllowTraffic.} Removes a specific drone IP address from the block list.

\item \textbf{Sleep.} Wait silently until this turn is over.

\end{itemize}

In addition to these actions, each drone can also explicitly broadcast a 16~bit message to all neighbouring drones within immediate communication range. It is possible to include the full message space in the action space, although this would result in an extremely large action space of $65592$ (i.e., $56 +2^{16}$).

\begin{description}
    \item[State Space.] Blue agents partially observe the state of the environment with decreasing perception as distance from the host increases. Locally, blue agents can observe whether the previous action was successful, the contents of the block list, the position of the drone and whether any malicious activity has been detected. Within their communication range, blue agents can also observe the location of neighbouring drones and how many sessions (agents) their host has. Full details of the state space provided by the standard AI gym wrapper are shown in Table~\ref{table:obs_space}.
\end{description}


\begin{table}[t]
\centering
\begin{tabular}{p{0.135\columnwidth}p{0.53\columnwidth}p{0.2\columnwidth}}
Index & Description & Values \\
\midrule
$0$ & Was the last action successful? & \makecell[l]{True: $0$ \\Unknown: $1$ \\False: $2$ } \\ \greyrule
$1:18$ & Block list. Index $i+1\rightarrow$ drone $i$ & \makecell[l]{Unblocked: $0$ \\Blocked: $1$ }\\\greyrule
$19$ & Malicious process on host & \makecell[l]{False: $0$ \\True: $1$ } \\ \greyrule
$20:38$ & \makecell[l]{Malicious network events list\\Index $20+i\rightarrow$ drone $i$} & \makecell[l]{None: $0$ \\1 event: $1$\\2 events: $2$ } \\ \greyrule
$39$ & Host $x$ position & \makecell[l]{$0\leq x\leq 100$}\\\greyrule
$40$ & Host $y$ position & \makecell[l]{$0\leq x\leq 100$}\\\greyrule
$41:108$ & For each neighbouring drone, a tuple $(id_i, x_i, y_i, s_i)$ specifying the drone ID $id_i$, coordinates $(x_i,y_i)$ and number of sessions $s_i$.& \makecell[l]{$0\leq x\leq 100$ } \\ \greyrule
$109:381$ & (Optional) A sequence of $16$~bit messages from neighbouring drones. & \makecell[l]{$\{0,1\}^{16\times 17}$ } \\
\bottomrule
\end{tabular}
\caption{The observation space provided to blue agents by the standard CAGE challenge gym wrapper.}
\label{table:obs_space}
\end{table}





\section{Baseline} \label{sec:baseline}
Before presenting our expert and MARL based agents for resilient communication networks, we motivate the need with a baseline evaluation of the CAGE challenge problem. In other words, 

\vspace{2.5ex}
\noindent\emph{How difficult is the CAGE challenge?}
\vspace{1.5ex}

\begin{description}

\item[Green Agent Noise.]
The reward function shown in Table~\ref{table:reward_function} only includes negatives, so the maximum possible score is trivially zero. However, because communicating green agents choose recipient drones at random, these choices include unroutable destinations and incur a negative reward. To improve the upper bound on maximum score we evaluated the green agent in the CAGE challenge scenario but with malware completely disabled. Over 1000 episodes, the standard evaluation size, green agents alone induce an average score of $-355.3$ with a standard deviation of $253.2$. 

\item[A Sleepy Defence.] The trivial worst-case score is $-9000$. To improve our bound on the minimum score we should deem interesting, we evaluate a team of blue agents who simply use the Sleep action. This provides no defence whatsoever and allows us to quantify the damage an unchecked malware is capable of. Over 1000 episodes, red and green agents with no defence result in an average score of $-8292.1$ with standard deviation $432.2$. 


\item[Basic Strategies] To further improve our bounds for a noteworthy defence, we evaluate the four basic expert agents included in the CAGE challenge source code. Once again we report the average score $\mu$ and standard deviation $\sigma$ over 1000 episodes.
\begin{itemize}
    \item \textbf{RemoveBlueDrone, $\mu = -7396.8,\sigma = 599.0$.}\\This agent always uses the RemoveOtherSessions action, trying to remove low-privilege red agents before they can escalate further. 
    
    \item \textbf{RetakeBlueDrone, $\mu = -8309.7,\sigma = 409.0$.} \\This agent always uses the RetakeControl action on a random target from the swarm. This will restore blue agents on some compromised drones. 
    
    \item \textbf{AdvancedRetakeBlueDrone, $\mu = -8087.4,\sigma = 592.8$.} \\Uses RetakeControl if malicious activity is detected on a neighbour, otherwise uses RemoveOtherSessions. This combines the previous two approaches and conserves bandwidth.

    \item \textbf{AdvancedBlockBlueDrone, $\mu = -8312.2,\sigma = 419.0$.}\\ As AdvancedRetakeBlueDrone except BlockTraffic is used on the target with suspected malicious activity instead of RetakeControl. 
    
\end{itemize}

\end{description}

\noindent
\textbf{Proximal Policy Optimisation (PPO)}
As our final baseline, we straightforwardly (i.e., with no changes to the environment) apply PPO to the CAGE challenge. PPO~\cite{schulman2017ppo} is an efficient policy gradient~\cite{sutton99pg} algorithm for DRL. PPO offers a comparatively simple implementation, helping to solve a broad range of difficult problems including competitive e-sports~\cite{openai2019dota}, emergent tool use~\cite{baker2020emergent} and microprocessor side-channel attack discovery~\cite{luo2023autocat}. In ACD, PPO has been used to successfully defend enterprise networks including top performances in the first two CAGE challenges~\cite{foley2022CAGE1,hannay2022cage2}. Furthermore, multi-agent PPO (MAPPO) achieves state-of-the-art performances in cooperative multi-agent environments~\cite{yu2022mappo}.

Taking standard parameters from the literature~\cite{andrychowicz2021whatmattersppo}, we use a fully connected neural architecture of width 256 and depth 2 with ReLU \cite{fukushima1975relu} activation between each layer. We train two models, first using the standard standard state space from Table~\ref{table:obs_space} and, second using the explicit 16~bit communication channel to broadcast the action selected by each agent\footnote{Using the \texttt{AgentCommsPettingZooParallelWrapper} included in CybORG}. The second experiment significantly increases the model size, increasing the state space from $109$ to $381$ and the action space from $56$ to $896$. We evaluate the best checkpoints from both models after training each for 25 million time steps. The average model scores over 1000 episodes, alongside the other baselines, are shown in Table~\ref{table:baseline_scores}.

\begin{table}[ht]
\centering
\begin{tabular}{p{0.44\columnwidth}cc}
Baseline method & Avg. reward & Std. dev.\\
\midrule
Sleep & $-8292.1$ & \makecell[l]{ $432.1$ } \\ \greyrule
RemoveBlueDrone & $-7396.8$ & \makecell[l]{ $599.0$ }  \\ \greyrule
RetakeBlueDrone & $-8309.7$ & \makecell[l]{ $409.0$ }  \\ \greyrule
AdvancedRetakeBlueDrone & $-8087.4$ & \makecell[l]{ $592.8$ }  \\ \greyrule
AdvancedBlockBlueDrone & $-8312.2$ & \makecell[l]{ $419.0$ }  \\ \greyrule
PPO & $-7617.8$ & \makecell[l]{ $651.3$ }  \\ \greyrule
PPO with explicit comms. & $-6745.3$ & \makecell[l]{ $945.4$ }  \\ \greyrule
\bottomrule
\end{tabular}
\caption{Baseline scores in the CAGE challenge.}
\label{table:baseline_scores}
\end{table}

\section{Canaries and Whistles}\label{sec:canary}
The results in the previous section highlight the importance of explicit communication and the favourable performance of even simplistic expert agents. When explicit communication is neglected, PPO fails to outperform a basic approach based on greedily preventing privilege escalation on the local host (i.e., the RemoveBlueDrone method). Building on this insight, and to provide a state-of-the-art benchmark for the CAGE challenge, we developed a new expert blue agent that utilises a system of canaries and whistles to actively mitigate drone malware. Our Canaries and Whistles (CW) agent, described in Algorithm~\ref{alg:canary}, is based on an explicit communication protocol that allows blue agents to keep track of neighbouring drones. Every time step that a host drone is not compromised, the blue agent broadcasts a canary message containing a unique identity number (UID). Meanwhile, blue agents also keep track of the canaries they receive at each step. Once a drone becomes compromised, its blue agent will be displaced by the red team and it will cease broadcasting canaries. Immediately, neighbouring drones still controlled by the blue team notice the missing Canary and apply a strategy to mitigate the malware. In addition, blue agents that detect a compromised neighbour become ``whistleblowers'' and broadcast the infected drone's UID alongside the host's own. Whistle messages are subsequently spread throughout the swarm by way of re-broadcasting. 

Our CW agent was designed by domain experts to tackle this challenge specifically. A major determining factor in the design is the 16~bit limit applied to explicit drone messages. However, this is enough to encode two drone UIDs and a bit to distinguish between original and re-broadcast whistles. Indeed, 5~bits remain unused and could be used to enhance the algorithm further. The \texttt{PAD} and \texttt{UNPAD} algorithms used to encode and decode messages, respectively, can be found in Appendix~\ref{appendix:cwdeps}. Over 1000 episodes, the CW agent averages a score of $-1577.7$ with a standard deviation of $800.4$. Based on the CAGE challenge competition phase~\cite{cage_challenge_3_announcement}, which is now closed, the CW agent provides a state-of-the-art performance.


\begin{algorithm}[]
\DontPrintSemicolon
\SetAlgoLined
\SetNoFillComment
\SetAlgoNoEnd
\SetKwInOut{Args}{Args}
\Args{$\text{d}_{\text{UID}}$, ${\text{position}}$}
$t\leftarrow -1$, $\text{neighbours}\leftarrow \emptyset$, $\text{to-fix}\leftarrow \emptyset$\;
\While{episode is not done}{
    \tcc{For each time step in the episode}
    $t\gets t+1$\;

    {\If{\text{this drone is infected}}{
        \tcc{Blow whistle on self}
        $\text{msg}\gets \texttt{PAD}(0,0,\text{d}_{\text{UID}})$\;
        $\text{action}\gets \text{RemoveOtherSessions}$\;
        \Return \text{msg}, \text{action}
    }
   }
    
  \eIf{$t>1$ and position has changed}{
    $\text{position}\gets\text{new position}$\;
    $\text{neighbours}\gets\emptyset$
  }{
    
        \For{ each $m\text{ in }\{\text{received messages}\}$}
        {
            \tcc{Parse explicit messages}
            $m_\text{canary}, m_\text{ovheard}, m_\text{whistle} \gets \texttt{UNPAD}(m)$\;
            $\text{neighbours}[m_\text{canary}] = t$\;
          
            \If{$\text{d}_{\text{UID}} \neq m_\text{canary} \land m_\text{ovheard}=1$}{
            \tcc{Notified of infected drone}
            \If{$\text{to-fix}[m_\text{whistle}]=\emptyset$}{
                $\text{to-fix}[m_\text{whistle}]\gets``\text{block}"$\;
            }
          }
        }
      
    }
    $\text{overheard}\gets 0$\;
      \For{$\text{each id in } \{\text{neighbours}\}$} 
      {
        \tcc{Check canaries}
        \If{$\text{neighbours}[\text{id}] = t-1$} {
            \tcc{Infected neighbour detected}
            $\text{to-fix}[\text{id}]\gets ``\text{block}"$\;
            $\text{whistle}\gets\text{id}$\;
            $\text{overheard}\gets 1$
        }
      }
    \If{$\text{to-fix} \neq \emptyset$}
    {
        \tcc{Try to repair neighbouring drone}
         $t_{ID}\xleftarrow{\$}\{$to-fix$\}$\;
         \If{to-fix$[t_{ID}] = ``\text{block}"$ }{
            rnd $\gets \{0.0\leq\mathbb{R}\leq 1.0\}$\;
            \eIf{rnd $\leq 0.225$}{
                $\text{to-fix}[\text{canary}]\gets ``\text{allow}"$\;
                $\text{action}\gets \texttt{RetakeControl}$\;
            }{  
                $\text{to-fix}[\text{canary}]\gets ``\text{retake}"$\;
                $\text{action}\gets \texttt{BlockTraffic}$\;
            }
         }
         \If{to-fix$[t_{ID}] = ``\text{retake}"$ }{
            $\text{to-fix}[\text{canary}]\gets ``\text{allow}"$\;
            $\text{action}\gets \texttt{RetakeControl}$\;
         }
    
         \If{to-fix$[t_{ID}] = ``\text{allow}"$ }{
            $\text{to-fix}[\text{canary}]\gets \emptyset$\;
            $\text{action}\gets \texttt{AllowTraffic}$\;
         }
         $\text{msg}\gets \texttt{PAD}(\text{d}_{\text{UID}},\text{overheard},\text{whistle})$\;
         \Return action, msg
     
    }
    \tcc{If no neighbours are infected then greedily prevent local privilege escalation}
    $\text{action}\gets \texttt{RemoveOtherSessions}$\;
        \Return action, msg
}

\caption{Canaries and Whistles Agent}
\label{alg:canary}
\end{algorithm}

\section{Joining the Opera}
Inspired by the performance of our CW agent we experimentally investigate the gap between straightforward PPO, PPO with explicit communication, and the CW agent. We identify three improvements that, when combined, outperform the CW agent alone and establish a new state-of-the-art result. Specifically 1) addressing limitations in the CybORG observation space, 2) a CL method determined by the finding that it's easier to learn from ``joining the opera'' (i.e., join the team of CW agents) than starting from scratch, and 3) defining a denoised reward function for optimised learning.

\subsection{Observation Space}
The CAGE challenge observation space provided by the CybORG environment is performance limiting when considering the behaviour of our expert CW agents. In particular, CW agents keep track of neighbouring drones over multiple time steps. This allows blue agents to keep track of which drones in the swarm could be compromised and respond accordingly. Considering the observation space in Table~\ref{table:obs_space}, it is not possible for the standard PPO agent we evaluated in Section~\ref{sec:baseline} to learn the CW strategy. Despite the ability of deep RL algorithms to tolerate POMDP environments~\cite{sutton2018rlbible}, they cannot parameterise non-Markovian variables. In other words, whether a drone is currently compromised (and needs blocking or retaking) is not independent of past observations given the present state. We address the limitations of the standard observation space with a new space, shown in Table~\ref{table:new_obs_space}, designed to incorporate the minimum features needed to learn a CW-like policy. We deliberately remove redundant variables (e.g., drone $x,y$ positions) to improve learning efficiency, and include new stateful variables that provide a more Markovian observation. The most significant changes include providing an estimation of whether a neighbouring drone needs fixing (based on CW messages which are processed transparently in the background) and reporting the last action performed on each specific neighbour. 

\begin{table}[b]
\centering
\begin{tabular}{p{0.135\columnwidth}p{0.53\columnwidth}p{0.2\columnwidth}}
Index & Description & Values \\
\midrule
$0$ & Host Drone UID & \makecell[l]{$\{0,\ldots,17\}$} \\ \greyrule
$1$ & Was the last action successful? & \makecell[l]{False: $0$ \\True: $1$} \\ \greyrule
$2$ & Last action number & \makecell[l]{$\{0,\ldots,55\}$} \\ \greyrule
$3:20$ & \makecell[l]{Last action type on neighbour\\Index $i+2\rightarrow$ drone $i$ UID}   & \makecell[l]{RetakeControl: $0$ \\BlockTraffic: $1$ \\ AllowTraffic $2$}\\\greyrule
$21$ & Malicious process on host & \makecell[l]{False: $0$ \\True: $1$ } \\ \greyrule
$22:39$ & Block list. Index $i+22\rightarrow$ drone $i$ & \makecell[l]{Unblocked: $0$ \\Blocked: $1$ }\\\greyrule
$40:57$ & \makecell[l]{Neighbour needs fixing\\Index $i+22\rightarrow$ drone $i$ UID} & \makecell[l]{False: $0$ \\True: $1$ }\\\greyrule
$58:330$ & (Optional) A sequence of $16$~bit messages from neighbouring drones. & \makecell[l]{$\{0,1\}^{16\times 17}$ } \\
\bottomrule
\end{tabular}
\caption{The revised observation space designed to allow learning CW-like policies.}
\label{table:new_obs_space}
\end{table}


\subsection{An Expert Curriculum}\label{subsec:curric}
CL is an approach that has been successfully used to master a number of MARL environments~\cite{wang2020coopcurric}, including in combination with PPO~\cite{baker2020emergent,yu2022mappo}. We identify a CL method, Opera, that allows for a gradual increase in the complexity of the CAGE challenge environment, helping to mitigate the difficulties (e.g., non-stationary learning and combinatorial explosion of the action space) associated with MARL environments. The essential observation is that, with a high-performing expert agent to hand, DRL agents can be incrementally introduced to the environment. We begin training just a single DRL agent, amongst a swarm of CW agents, and then gradually increase the number of learning-based agents until maximum performance is reached. For this purpose we create a series of wrappers for the CybORG environment that allow us to specify an arbitrary mixture of CW and PPO agents. To prevent bias in our models we ensure they are hosted randomly every episode, placing different agents on different drone hosts. 

%
\subsection{A denoised reward}\label{subsec:denoise}
The CAGE challenge scoring function includes a negative penalty for communications that are unroutable. As discussed in Section~\ref{sec:baseline}, this stochastically decreases the maximum agent score based on the ad-hoc network topology and randomised choosing of message recipients. Unfortunately, this directly antagonises DRL-based approaches which are optimised exclusively on the maximisation of cumulative reward. This is further exacerbated by the partial observability of each agent, which additionally obfuscates the relationship between state, action and reward. To train our new agents we modify the CybORG environment to distinguish between messages that are unroutable and those that are blocked, dropped or intercepted. We then remove the negative reward for unroutable messages during training (i.e., we still evaluate our models using the reward in Table~\ref{table:reward_function}), providing improved learning. 

\section{Evaluation} \label{sec:eval}
Here we evaluate the performance of our learning-based PPO policies, investigate our proposed improvements to the observation space, scrutinise our mixed Opera method of CL (i.e., combining CW and PPO agents) and compare the behaviour of our learning-based agents with their expert collaborators. Our best performing model overall utilises a 7:11 split of CW:PPO agents. The learning-based agents are trained using an expert curriculum ranging from 1:17 to 14:4. We also use our improved observation space from Table~\ref{table:new_obs_space} and the denoised reward strategy. Averaged over 1000 episodes, our best mixed Opera scores $-1487.9$ with a standard deviation of $626.1$. 

\subsection{Observation Space}
\begin{figure}[t]
    \centering
    \includegraphics[width=1.0\columnwidth]{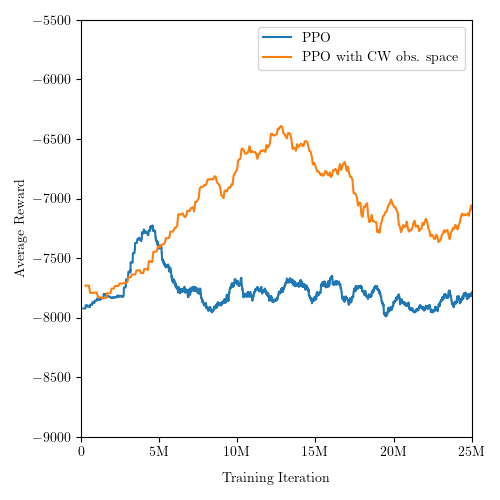}
    \caption{Comparing the learning performance of baseline PPO with and without our improved observation space.}
    \label{fig:learn_obsspace}
\end{figure}

Our proposed observation space allows learning-based agents to keep track of whether neighbouring drones are likely to be infected with malware. In Figure~\ref{fig:learn_obsspace}, we compare the learning curves of the baseline PPO result from Section~\ref{sec:baseline} and our new observation space from Table~\ref{table:new_obs_space}. In both cases we train a single policy (with width 256 and depth 2) for all 18 agents simultaneously. Each policy took approximately 3.5 hours to train on a single machine with a 24-core Intel i9-13900K, 64 GB of memory and a 24GB RTX 4090 GPU. Table~\ref{table:obs_scores} shows the corresponding policy scores, indicating that our changes do indeed improve performance. We believe that significant further score improvements in this setting (i.e., training 18 learning-based agents simultaneously) are unlikely without utilising methods for centralised coordination such as centralised training decentralised execution (CTDE)~\cite{yu2022mappo}.


\begin{table}[h]
\centering
\begin{tabular}{p{0.3\columnwidth}cc}
Observation Space & Avg. reward & Std. dev.\\
\midrule
As in Table~\ref{table:obs_space} & $-7617.8$ & \makecell[l]{ $651.3$ }  \\ \greyrule
As in Table~\ref{table:new_obs_space} & $-6884.4$ & \makecell[l]{ $845.4$ }  \\
\bottomrule
\end{tabular}
\caption{Observation-space related score changes.}
\label{table:obs_scores}
\end{table}

\begin{figure}[!hb]

    \centering
    \begin{subfigure}{0.97\columnwidth}
    \includegraphics[width=0.98\columnwidth]{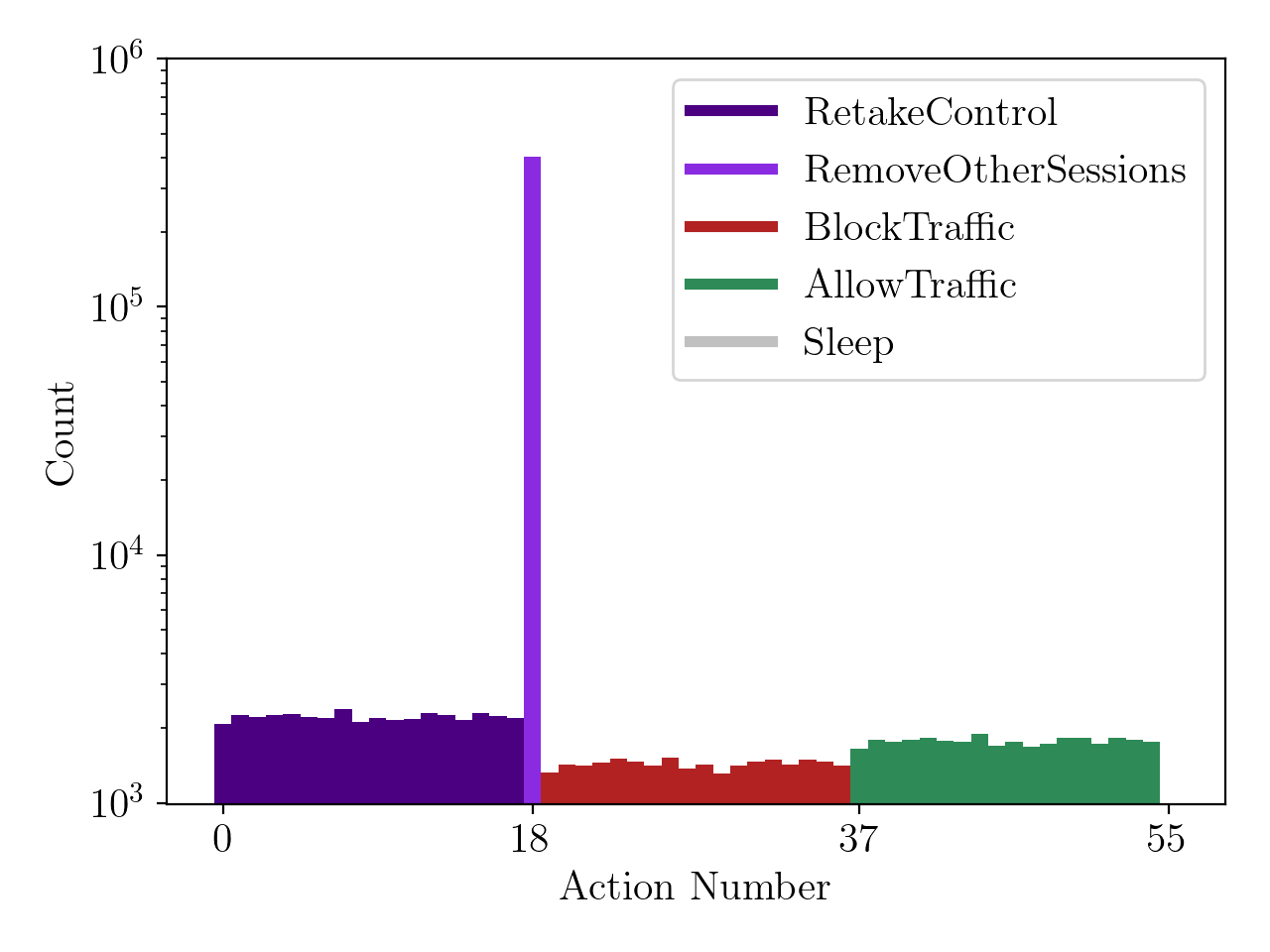}
    \caption{Expert CW agent.}
    \end{subfigure}

    \begin{subfigure}{0.97\columnwidth}
    \includegraphics[width=0.98\columnwidth]{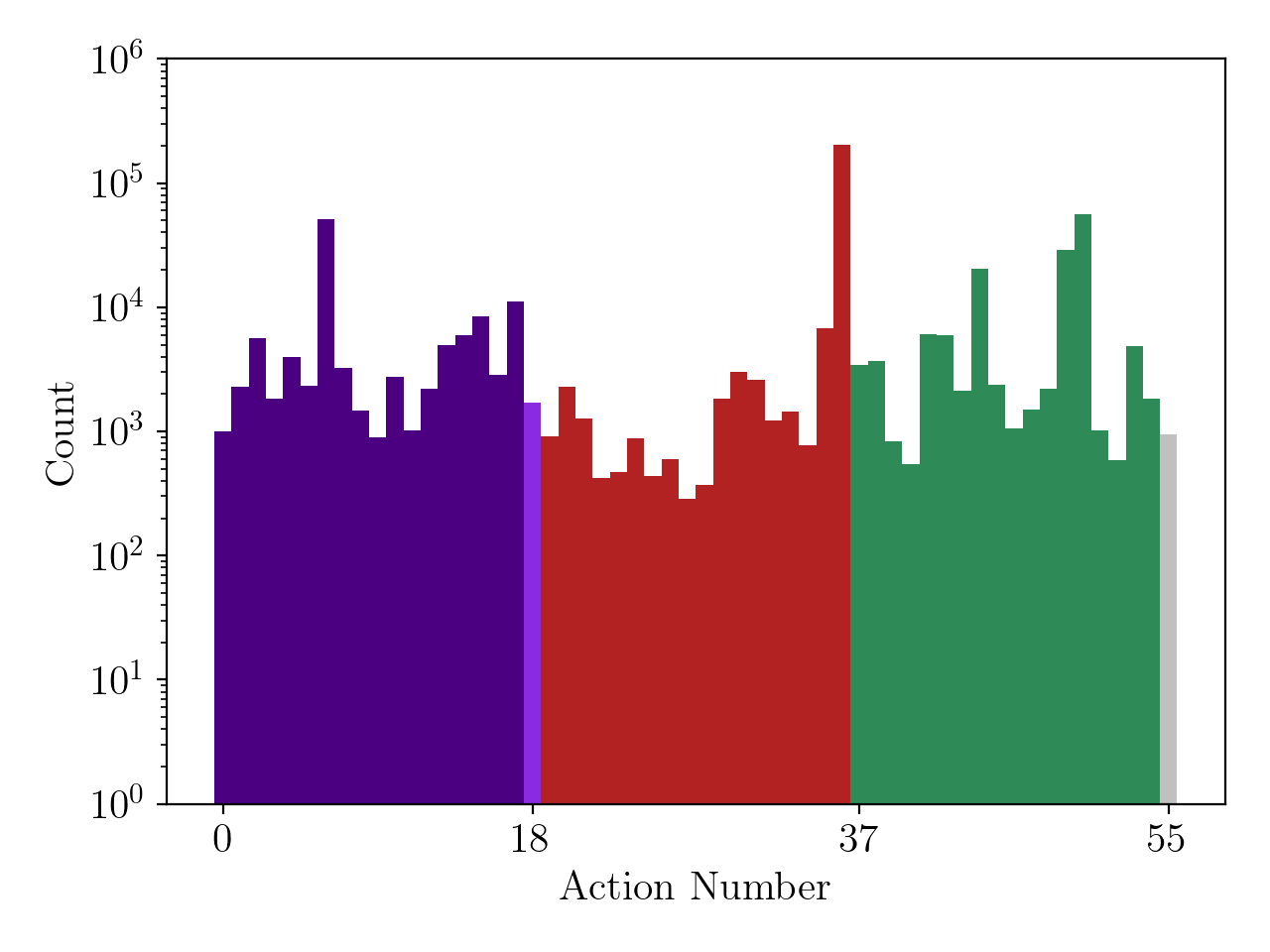}
    \caption{Baseline PPO agent.}
    \end{subfigure}

    \begin{subfigure}{0.97\columnwidth}
    \includegraphics[width=0.98\columnwidth]{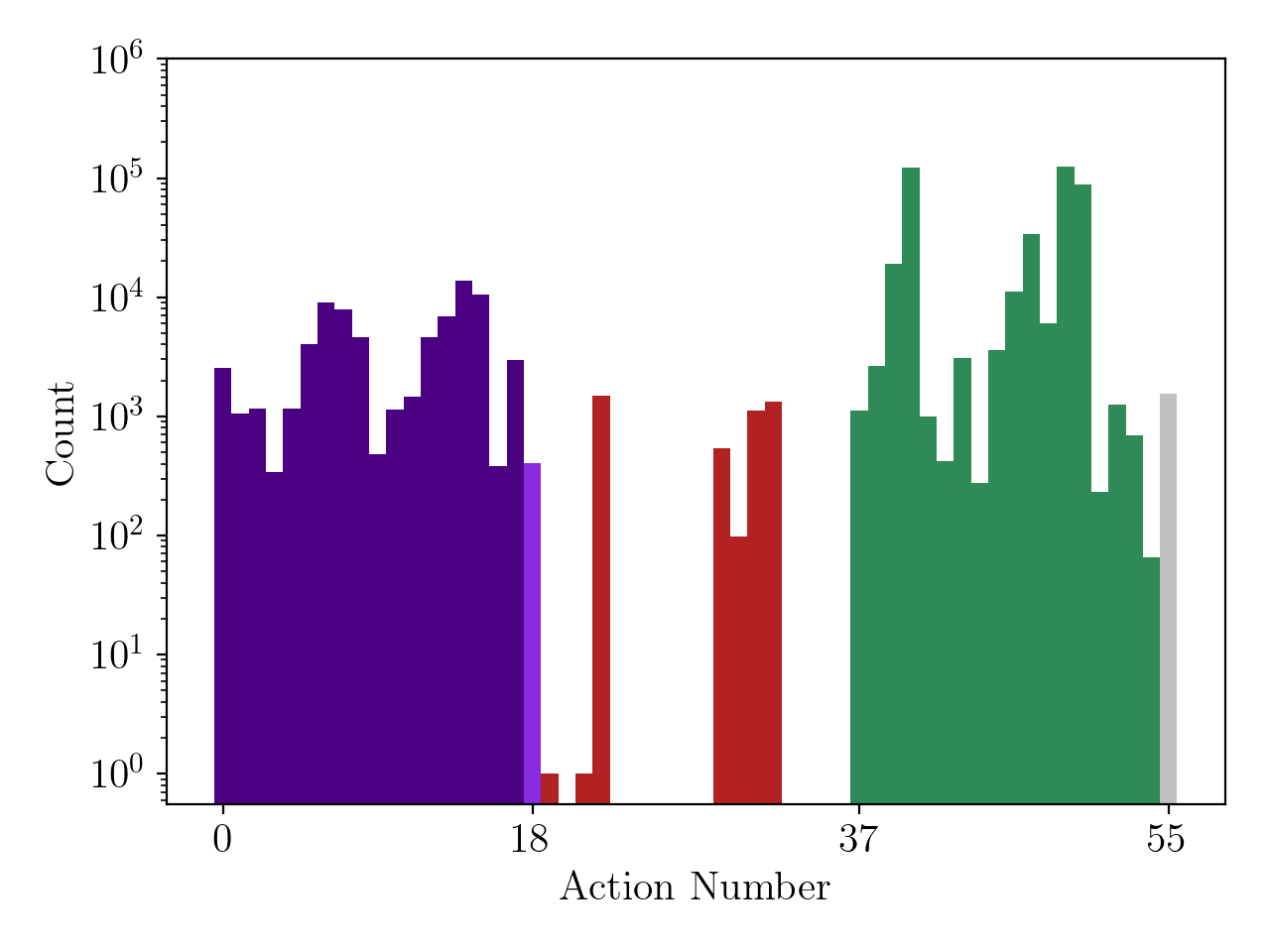}
    \caption{Mixed Opera PPO agent.}
    \end{subfigure}
    
\caption{Action distributions for the CW, baseline PPO and mixed Opera PPO agents. For normalisation, actions are sampled from a single agent chosen randomly each episode.}
\label{fig:policy_hists}
\end{figure}






\subsection{Mixed Opera}
\begin{figure}[ht]
    \centering
    \includegraphics[width=1.0\columnwidth]{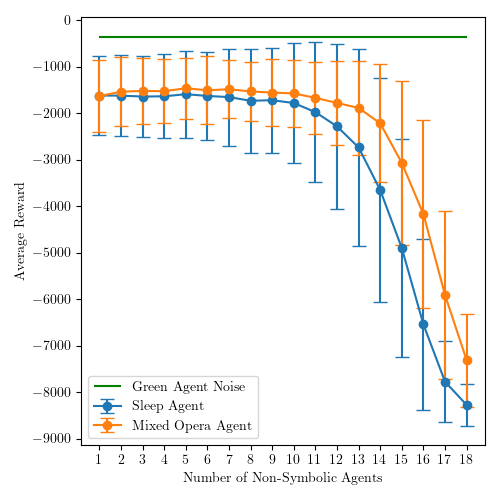}
    \caption{The change in average score and standard deviation as the number of learning-based agents is increased from 1/18 to 18/18.}
    \label{fig:compare1hot}
\end{figure}

Our Opera method incrementally introduces learning-based agents to a population of expert influencers until they learn all of the skills they need (or are capable of). We evaluate our approach in Figure~\ref{fig:compare1hot} by calculating the average score as CW agents are incrementally substituted for their learning-based contemporaries. The policy is trained iteratively using the expert curriculum method, beginning with 1 agent and gradually increasing the number to 14. Training was stopped at this point as beyond 14 performance declined rather than improved. As a baseline, we also substitute an inactive agent that takes no actions whatsoever (i.e., the Sleep Agent). 

\begin{table}[h]
\centering
\begin{tabular}{p{0.3\columnwidth}cc}
Reward & Avg. reward & Std. dev.\\
\midrule
Noisy & $-6770.5$ & \makecell[l]{ $1064.0$ }  \\ \greyrule
Denoised & $-6884.4$ & \makecell[l]{ $845.4$ }  \\
\bottomrule
\end{tabular}
\caption{Denoised reward related score changes.}
\label{table:denoised-reward}
\vspace{-3ex}
\end{table}


\subsection{Policy Analysis}
As shown in Figure~\ref{fig:policy_hists}, we investigate the coarse-grained policy differences between the expert CW agent, the baseline PPO agent and our mixed Opera agent by counting the actions chosen over 1000 randomised episodes. The CW agent distinctively chooses RemoveOtherSessions, locally removing low-privilege red agents placed by neighbouring infected drones, just over 80\% of the time. The remaining actions are distributed more evenly between RetakeControl, BlockTraffic and AllowTraffic. As the red team's trojan maliciously blocks traffic to disrupt communication, AllowTraffic is chosen more often than BlockTraffic. In comparison, the baseline PPO agent chooses from the available actions (including the sleep action which does nothing) with a more equal probability.






\section{Discussion} \label{sec:commentary}
Compared with previous ACD gyms, the third CAGE challenge represents a substantially more challenging learning environment. This is evident, in addition to the results presented in this paper, from the challenge leaderboard which currently scores expert agents from multiple teams significantly higher than any of the learning-based agents. Nevertheless, multi-agent algorithms are well motivated for ACD because they can potentially divide and conquer impractically large action spaces into more manageable sub tasks attended by specialised agents. It is remarkable however, that the CAGE challenge has a large number of agents relative to the wider literature where $20$ is usually the maximum (e.g., \cite{singh2018individualized, wang2019varagents}) and 2-6 is far more common (e.g.,~\cite{baker2020emergent}). The only other MARL ACD environment currently supports a maximum of two agents~\cite{kunz2022multiagent} trained competitively. Regarding our approaches to MARL in the CybORG environment, it is notable that we have not applied any approaches based on a centralised critic. The success of PPO in multi-agent environments has generally otherwise been based on the CTDE paradigm~\cite{yu2022mappo}.

As in the previous two CAGE challenges, the issue of determining the optimal score is a difficult one. There currently is no principled approach to knowing how close various approaches are, leaving the possibility that significant performance improvements are yet to be made. Certainly, although it can be understood as resulting from the impact of drone malware, there is a gap between the ceiling generated by green agents (as discussed in Section~\ref{sec:baseline}) and the state-of-the-art presented here. Finally, we note that the design of the CybORG gym could be improved regarding parallelisation. A major bottleneck in the training and evaluation of our agents is the need to wait for a single thread to simulate the environment before returning new observations and rewards to each agent. Multi-agent AI gyms should ideally be designed with parallelisation to match the algorithms that will train agents within them.





\section{Related Work}

To the best of the authors knowledge no other publications have applied MARL to the problem of resilient ad-hoc communications in the presence of malware infected hosts~\cite{nguyen2021drl4sec}. Indeed, there is only one other open source MARL environment for ACD beyond CybORG~\cite{kunz2022multiagent, vyas2023acd} and both were released only in the last 12 months. 

Concerning single-agent RL for ACD, the literature is considerably more established. Foley et al.~\cite{foley2022CAGE1} showed the effectiveness of hierarchical PPO, based on choosing a specialised sub-agent every time step, in tackling two differentiated red agents in the first CAGE challenge. In later work, Foley et al~\cite{foley2022CAGE2} further determine that the hierarchical DRL supervisor can be replaced with a more effective bandit-based algorithm for choosing the best agent. In the same work, the authors also study the explainability of DRL agents for ACD, apply an ablation study, and calculate SHapley Additive exPlanations (SHAP) values~\cite{lundberg2017shap} to determine the importance of features in the CybORG environment. There are several recent surveys of autonomous and automated cyber defence~\cite{nguyen2021drl4sec, adawadkar2022rl4sec, huang2022resilientcyber, sewak2022drl4sec} including by Vyas et al.~\cite{vyas2023acd} who comprehensively survey the current landscape of autonomous cyber operations gyms and develop a set of requirements they use to evaluate each environment. Beyond CybORG~\cite{cage_cyborg_2022}, CyberBattleSim~\cite{msft2021cyberbs} and Yawning Titan (YT)~\cite{dhir2022causalyt} are notable for also providing open source simulation based environments for attacking and defending computer networks, respectively. Beyond ACD, the RL paradigm is increasingly accomplishing remarkable results in a range of systems security environments including evading hardware trojan detection~\cite{gohil2022attrition}, finding new injection attacks~\cite{wahabi2023sqirl, foley_haxss_2022}, closed-box malware generation~\cite{song2022mabmalware}, and microprocessor cache-timing vulnerability discovery~\cite{luo2023autocat}. 

Related to our mixed Opera method of CL, Campbell et al.~\cite{campbell2023curricrl4sec} propose a curriculum framework for autonomous network defense using MARL. Yu et al.~\cite{yu2022mappo} study the performance of multi-agent PPO in cooperative environments and discover surprisingly strong performance that is competitive to conventional off-policy methods. Piterbarg et al.~\cite{piterbarg2023nethack} study the NeurIPS 2021 NetHack Challenge and also discover a gap between expert and learning-based approaches. The authors use different methods based on a hierarchical action space, neural architecture improvements and imitation learning to bridge the gap for their agents.

\section{Conclusion} \label{sec:conclusion}
In this paper we present the full details of our state-of-the-art expert CW agent for decentralised autonomous malware resilience in an ad-hoc drone network. We use CW to address some of the deficiencies in prior learning-based agents, identifying three specific methods for doing so: (1) Providing a (more) Markovian observation space, (2) implementing CL by gradually increasing the proportion of learning-based agents, and (3) creating a noise-free reward function, allowing us to considerably close the gap in performance between expert and learning-based agents. Finally, we present a new state-of-the-art result in the third CAGE challenge based on a mixed Opera of expert and learning-based Canary agents. 

In future work we will consider emergent communication and whether protocols similar to CW could be learned automatically without the need for an explicit specification. In addition, since red agents will inevitably try to use our protocols against us, exploring communication learned (e.g., using autocurricula~\cite{baker2020emergent}) under competitive adversarial pressure is a compelling direction. More broadly in ACD, many open problems remain including improved methods for coordinating shared situational awareness in decentralised agents; scalable observation spaces able to deal with an arbitrary number of neighbouring drones, algorithms for lifelong learning, and robustness for autonomous agents.

\begin{acks}
Research funded by the Defence Science and Technology Laboratory (Dstl) which is an executive agency of the UK Ministry of Defence providing world class expertise and delivering cutting-edge science and technology for the benefit of the nation and allies. The research supports the Autonomous Resilient Cyber Defence (ARCD) project within the Dstl Cyber Defence Enhancement programme.
\end{acks}

\bibliographystyle{ACM-Reference-Format}
\bibliography{bibliography}
\ \\

\appendix

\section*{Appendix}
Supporting information.

\section{CW Agent Dependencies}\label{appendix:cwdeps}
The CW agent algorithm (Algorithm~\ref{alg:canary}) depends on two subroutines to encode and decode explicit communication messages. These are provided by Algorithm~\ref{alg:pad16} and Algorithm~\ref{alg:unpad16}, respectively. 

\vspace{4ex}
{
\removelatexerror
\begin{algorithm}[h]
\DontPrintSemicolon
\LinesNumbered
\SetAlgoLined
\SetNoFillComment
\SetKwInOut{Args}{Args}
\Args{$\text{canary}, \text{overheard}, \text{whistle}$}
\tcc{Pad canary, whistle and overheard into a 16~bit binary string}
Convert canary to binary and pad to 5~bits.\;
Convert overheard to 1~bit binary.\;
Convert whistle to binary and pad to 5~bits.\;
\Return $(\text{whistle}<<11) \lor (\text{overheard}<<6) \lor \text{canary}$
 \caption{\texttt{PAD}}\label{alg:pad16}
\end{algorithm}
}

\vspace{0ex}
{
\removelatexerror
\begin{algorithm}[h]
\DontPrintSemicolon
\LinesNumbered
\SetAlgoLined
\SetNoFillComment
\SetKwInOut{Args}{Args}
\Args{$m\in\{0,1\}^{16}$}
\tcc{Recover the canary, whistle and overheard values from a 16~bit binary string}
$\text{canary}\gets m\land\mathtt{0x3F}$\;
$\text{overheard}\gets (m>>6)\land\mathtt{0x1}$\;
$\text{whistle}\gets (m>>11)\land\mathtt{0x3F}$\;
\Return (canary, overheard, whistle)
 \caption{\texttt{UNPAD}}\label{alg:unpad16}
\end{algorithm}
}

\end{document}